\documentclass[aps,twocolumn,superscriptaddress,amsmath,amssymb,nofootinbib]{revtex4-2}
\usepackage{graphicx,enumitem}
\usepackage{standalone}
\usepackage{multirow,dcolumn}
\usepackage{txfonts}
\usepackage{hyperref}
\usepackage{soul,cancel}
\usepackage{epsf,amsmath,amssymb,verbatim,color,multirow,pifont,graphicx,cleveref,tabularx}
\usepackage[dvipsnames]{xcolor}
\usepackage{rotating}
\usepackage{amsmath,amsfonts,amssymb,bm,cancel}
\newcommand{\paulix}[1]{{\sigma}_{#1}^{x}}
\newcommand{\pauliy}[1]{{\sigma}_{#1}^{y}}
\newcommand{\pauliz}[1]{{\sigma}_{#1}^{z}}
\newcommand{\n}[1]{{n}_{#1}}
\newcommand{\pauliplus}[1]{{\sigma}_{#1}^{+}}
\newcommand{\pauliminus}[1]{{\sigma}_{#1}^{-}}
\input{Qcircuit}

\usepackage{tikz}
\usetikzlibrary{matrix}
\usepackage{braket}

\begin{document}
\title{
Simulating open quantum many-body systems using optimised circuits in digital quantum simulation}
	
\author{Minjae Jo}
\email{m.jo@imperial.ac.uk}
\affiliation{QOLS, Blackett Laboratory, Imperial College London SW7 2AZ, UK}

\author{M. S. Kim}
\affiliation{QOLS, Blackett Laboratory, Imperial College London SW7 2AZ, UK}

	
\begin{abstract}
Digital quantum computers are potentially an ideal platform for simulating open quantum many-body systems beyond the digital classical computers. Many studies have focused on obtaining the ground state by simulating time dynamics or variational approaches of closed quantum systems. However, dynamics of open quantum systems has not been given much attention with a reason being their non-unitary dynamics not natural to simulate on a set of unitary gate operations in quantum computing. Here we study prototypical models in open quantum systems with Trotterisations for the modified stochastic Schr{\"o}dinger equation (MSSE). Minimising the leading error in MSSE enables to optimise the quantum circuits, and we run the optimised circuits with the noiseless \textit{quantum assembly language (QASM) simulator} and the noisy IBM Quantum devices. The \textit{QASM simulator} enables to study the reachable system size that is comparable to the limits of classical computers. The results show that the nonequilibrium critical phenomena in open quantum systems are successfully obtained with high precision. Furthermore, we run the algorithm on the IBM Quantum devices, showing that the current machine is challenging, to give quantitatively accurate time dynamics due to the noise. Despite errors, the results by IBM devices qualitatively follow the trend of critical behaviour and include a possibility to demonstrate quantum advantage when the noise is reduced. We discuss how much noise should be reduced for a certain fidelity using the noise model, which will be crucial to demonstrate quantum advantage from future quantum devices. 
\end{abstract}
	
\maketitle
	
	


Digital quantum simulation (DQS) has offered one of the most promising applications of near-term quantum computers~\cite{blatt2012, monroe2021}. Contrary to analogue quantum simulation, which involves the construction of engineered quantum system emulating the target Hamiltonian, DQS involves designing a time evolution with a sequence of unitary gates to approximate unitary evolution. 
A recent experimental progress of DQS has been made with applications in condensed matter physics~\cite{lanyon2011, smith2019}, quantum chemistry~\cite{lanyon2010}, and high-energy physics~\cite{bermudez2010}. 
Those of DQS algorithms focus on finding the ground state of a quantum system, using the quantum phase estimation~\cite{cruz2020}, and variational quantum eigensolver~\cite{mcclean2018,bravo2020}, and quantum imaginary time evolution~\cite{motta2020}. Although the algorithms of a closed quantum system have been widely studied, fewer studies have examined the algorithms of open quantum systems because non-unitary dynamics is not natural to simulate on a unitary quantum hardware. 

On the other hand, recent advances using classical computers show that dynamics of an open quantum system exhibits novel physical phenomena originating from the interplay between coherent and incoherent dynamics. 
Examples include time crystals~\cite{gambetta2019}, driven-dissipative strong correlations~\cite{tomita2017,ma2019}, and dissipative phase transitions in the nonequilibrium steady-state~\cite{sieberer2013,diehl2010,torre2010,tauber2014,marino2016}. 
In particular, we focus on the critical phenomena of dissipative phase transitions, which is characterised by the concept of the universality class defined by a set of values of critical exponents. 
According to the universality, critical behaviour associated with continuous phase transitions has universal features independent of microscopic details of the system and can be classified into a small number of universality classes~\cite{altland2010}. Identifying the universality class is therefore a relevant task because it provides a classification scheme.
For the open quantum many-body systems, there arises a possibility that the interplay between coherent and incoherent dynamics produces the novel type of universality class that was not observed in classical models~\cite{lang2016,marino2016}.
For instance, the universality class of quantum contact process (QCP) changes from the classical directed percolation class~\cite{cardy1980} to quantum directed percolation class~\cite{carollo2019, jo2021}. 

In general, the computational resources required to simulate quantum systems using classical computers scale exponentially with the allowing degrees of freedom~\cite{kohn1999}. 
As a result, complete numerical descriptions of the general quantum many-body problems, where reduction schemes such as tensor network method for the Hilbert space of the system are impossible or unknown, can only be achieved for small systems by classical computers. 
This poses a challenge because many-body problems of condensed matter physics are directly governed by the system size, which needs to be large to describe the physics of the system accurately. 
In this respect, DQS is attractive to obtain quantum advantage by performing the large system size simulation that is intractable by classical computers~\cite{lloyd1996}. 

Recently, algorithms have been suggested to obtain this goal based on the Kraus decomposition~\cite{perez2020,re2020} and variational approaches~\cite{endo2020,yoshioka2020,haug2020}, and application of quantum imaginary time evolution to the open quantum system~\cite{kamakari2021}. Despite many advances, these algorithms also have potential disadvantages particularly to deal with the large system size. 
For example, even though the Kraus representation is efficient for the well-known noisy channel, obtaining the Kraus operator from the general Lindblad equation with large system size is challenging. Variational approaches offer an attractive alternative for simulating open system dynamics, but as in the case of closed systems they require an ansatz and a high dimensional classical optimization with large system size leading to an NP-hard problem~\cite{bittel2021}. Also, the application of quantum imaginary time evolution requires that the number of linear equations scale exponentially with the relevant qubit domain~\cite{gomes2020}.


Here, we simulate the Markovian open quantum system with a large system size using the Trotterised quantum simulation. To design the quantum circuits describing the dynamics of open quantum systems, we modify the stochastic Schr{\"o}dinger equation which unravels the evolution of the density matrix into trajectories of pure states. By minimising the leading error in the modified stochastic Schr{\"o}dinger equation (MSSE), we obtain the optimised quantum circuits showing the accurate dynamic results at a relatively large discretised time interval $\Delta t$. Further, we use a single ancilla qubit with the reset gate and mid-circuit measurements to describe the non-unitary evolution. Minimising the ancilla qubit is possible for the complex Lindblad operator which involves a pair of nearest-neighbours, and we exploit the DQS algorithm to perform the simulation of large system size comparable to the limits of classical algorithm. 

We apply the DQS algorithm on the prototypical models in open quantum systems including the dissipative transverse Ising (DTI) model~\cite{ates2012,jin2018,rose2016,hu2013} and the QCP model~\cite{marcuzzi2016, jo2019, gillman2021, gillman2022}. Firstly, the DQS algorithm is tested on the DTI model with the comparison between the optimised circuit and the original circuit. Then, we consider the QCP model showing the nonequilibrium phase transition in open quantum systems, which requires the large system size to measure critical exponents accurately.
For the QCP model, we note that the tensor network method fails when obtaining the critical exponent associated with the spatial correlation~\cite{carollo2019}. Thus, it was studied by the exact simulation, called the quantum trajectory method (also known as the quantum jump Monte Carlo simulation)~\cite{pleino1998} whose maximum system size $N$ would be $20$.
We use the \textit{QASM simulator}, and demonstrate that the nonequilibrium phase transition of the QCP model can be reproduced with the system size $N=26$.


Next, we use the IBM Quantum devices that support the reset gate and mid-circuit measurements. In order to implement the DQS algorithm, the connectivity is crucial because it requires a single ancilla qubit connected to the rest of the system qubits; however, a limitation of current IBM Quantum devices is the connectivity between qubits available in hardware. The connectivity issue requires more SWAP gates (each SWAP gate is decomposed by three CNOT gates) resulting in serious two-qubit gate errors. A workaround is feasible with \textit{ibmq\_tokyo} as it produces relatively high connectivity.
For the small system size, we also use \textit{ibm\_hanoi}. As the system size increases, however, the usefulness of IBM Quantum devices is limited by the noise leading to unreliable data.

To characterise the impact of the noise, we explore the noise model on the IBM Quantum devices. In particular, we focus on the single- and two-qubit gate errors modelled by the depolarising channel which describes average noise in real devices for a large circuit~\cite{urbanek2021,vovrosh2021}. We demonstrate that the data of our noise model describe those of the IBM Quantum device with high fidelity. Further, we discuss how much errors should be reduced to achieve certain fidelity, and how large the system can be simulated depending on the error levels.

The remainder of this paper is organized as follows. First, we present the modified stochastic Schr{\"o}dinger equation to design the quantum circuit describing the dynamics of open quantum systems. Next, two prototypical models in open quantum systems are introduced to demonstrate the DQS algorithm by the \textit{QASM simulator} and IBM Quantum devices. Furthermore, we investigate the noise model by controlling the gate noise and discuss how many qubits can be simulated at fixed gate noise. Finally, we summarise and draw our conclusions.



\section*{Modified stochastic Schr{\"o}dinger equation}
Here, we study the dynamics of open quantum systems and modify the relevant equations. For convenience, we take $\hbar=1$.
Time evolution of the open quantum system is described by the Lindblad equation, which consists of the Hamiltonian and dissipative terms:
\begin{align}
	\label{eq:lindeq}
	\partial_t{\rho}&=-i\left[ {H},{\rho} \right]
	+ \sum_{\ell=1}^N\left[ {L}_{\ell}{\rho} {L}^{\dagger}_{\ell}
	-\frac{1}{2} \left\{ {L}^{\dagger}_{\ell}{L}_{\ell},{\rho} \right\} \right]\,,
\end{align}
where ${\rho}$ is the density matrix of the full system and ${H}$ is the system Hamiltonian, and ${L}_\ell$ is the Lindblad operator at a site $\ell$. 
Simulating the density matrix in Eq.~\eqref{eq:lindeq} is not easy for a large system size because it requires a large number of qubits. Thus, instead of dealing with the density matrix, we consider the stochastic Schr{\"o}dinger equation~\cite{jacobs2014, wiseman2009}
\begin{align}
	d\ket{\psi(t)} &= \left[
	\left(-i{H}-\frac{1}{2}\sum_\ell\left({L}^{\dagger}_\ell{L}_\ell-\braket{{L}_\ell^\dagger {L}_\ell}\right)\right) dt \right.\nonumber\\
	&\left.
	+\sum_\ell \left( \frac{{L}_\ell}{\sqrt{\braket{{L}_\ell^\dagger {L}_\ell}}}-1\right)dN_\ell(t)
\right] \ket{\psi(t)}\,,
\label{eq:SSE}
\end{align}
where the increment $dN_\ell$ are mutually independent increments of Poisson noise. The probabilities for $dN_\ell=1$ and $dN_\ell=0$ for all $\ell$ during the time step $dt$ are given as
\begin{align}
	P(dN_\ell=1)=\braket{L^\dagger_\ell L_\ell}dt,\quad
	P(dN_\ell=0)=1-\sum_\ell \braket{L^\dagger_\ell L_\ell}dt.
	\label{eq:prob_poisson}
\end{align}
By taking the ensemble average of Eq.~\eqref{eq:SSE} ${\rho}(t)=E[ \ket{\psi(t)}\bra{\psi(t)}]$, the Lindblad master equation in Eq.~\eqref{eq:lindeq} is obtained~\cite{daley2014, breuer2002}. 
Inspired by the quantum trajectory method~\cite{pleino1998}, we manipulate this stochastic Schr{\"o}dinger equation to design the quantum circuits. To this end, we split the Hamiltonian and Lindblad evolution, which is given by
\begin{align}
	\ket{\psi(t+dt)} &= e^{-i(1-x)H dt}\left[
	1-\frac{1}{2}\sum_\ell\left({L}^{\dagger}_\ell{L}_\ell-\braket{{L}_\ell^\dagger {L}_\ell}\right) dt \right.\nonumber\\
	&\left.
	+\sum_\ell \left(\frac{{L}_\ell}{\sqrt{\braket{{L}_\ell^\dagger {L}_\ell}}}-1\right)dN_\ell(t)
\right] e^{-ixHdt} \ket{\psi(t)}\,,
\label{eq:SSE_2}
\end{align}
with a variable $x$ originating from non-commutativity between the Hamiltonian and Lindblad operators. 
Note that the It{\^o} rule for Poisson processes leads to $dN_\ell(t)dt=0$~\cite{jacobs2014,keys2020} and the terms up to $\mathcal{O}(dt)$ in Eq.~\eqref{eq:SSE} and Eq.~\eqref{eq:SSE_2} are the same.
Then, Eq.~\eqref{eq:SSE_2} can be further simplified as
\begin{align}
	\ket{\psi(t+dt)} &= e^{-i(1-x)Hdt}\left[
	\sum_\ell dN_\ell\frac{{L}_\ell}{\sqrt{\braket{{L}_\ell^\dagger {L}_\ell}}}+
	\left( 1-\sum_\ell dN_\ell \right)   \right. \nonumber\\
	&\left.
	\times \left( 1-\frac{1}{2}\sum_\ell\left({L}^{\dagger}_\ell{L}_\ell-\braket{{L}_\ell^\dagger {L}_\ell}\right) dt \right) 
\right]e^{-ixHdt} \ket{\psi(t)}\,.
\label{eq:SSE_3}
\end{align}
The Lindblad jump operation in square brackets of Eq.~\eqref{eq:SSE_3} can be interpreted as the system undergoes the dynamics that yields two possible outcomes. With probability $P(dN_\ell=1)$ in Eq.~\eqref{eq:prob_poisson}, the system jumps by one of the Lindblad operators ${L}_\ell$
\begin{align*}
\ket{\psi}\to \frac{{L}_\ell e^{-ixHdt} \ket{\psi}}{\sqrt{\braket{{L}_\ell^\dagger {L}_\ell}}}\equiv \ket{\psi_{\rm{jump},\ell}}\,.
\end{align*}
Otherwise, with probability $P(dN_\ell=0)$ in Eq.~\eqref{eq:prob_poisson}, the non-unitary evolution occurs 
\begin{align*}
\ket{\psi}\to \frac{\left(1-\sum_\ell\frac{1}{2}{L}^{\dagger}_\ell{L}_\ell dt\right)e^{-ixHdt} \ket{\psi}}{\sqrt{1-\sum_\ell\braket{{L}^{\dagger}_\ell {L}_\ell}dt}}\equiv \ket{\psi_{\rm evol}}\,.
\end{align*}
Under this interpretation, we consider the ancilla qubits $\ket{a}$ whose number equals to the number of Lindblad operators. However, the number of ancilla qubits can be reduced by one if we use the reset gate and mid-circuit measurement that are supported by IBM Quantum devices. Then Eq.~\eqref{eq:SSE_3} can be written as
\begin{align}
	\ket{\psi(t+dt)}\otimes\ket{a} &= e^{-i(1-x)Hdt}\left[
\sum_\ell \sqrt{P(dN_\ell=1)} \left(\ket{\psi_{\rm{jump},\ell}}\otimes\ket{1}\right) \right. \nonumber\\&
\left. +\sqrt{P(dN_\ell=0)} \left( \ket{\psi_{\rm evol}}\otimes \ket{0}\right) 
\right] \,.
\label{eq:MSSE}
\end{align}
The MSSE of Eq.~\eqref{eq:MSSE} is divided into unitary Hamiltonian evolution and non-unitary Lindblad evolution to design the quantum circuits appropriately. Note that the quantum trajectory method is different from the MSSE for any $x$ in that Hamiltonian evolution is included by non-unitary evolution.
Next, the optimum sequence of $x(t)$ can be determined by the norm of the next leading order $(dt)^2$ in Eq.~\eqref{eq:SSE_2}, which is given by
\begin{align} 
	&\Bigg|\Bigg|\bigg\{ -\frac{1-2x+2x^2}{2}H^2+\frac{i}{2}H\,\bigg(\sum_\ell L_\ell^\dagger L_\ell-\langle L_\ell^\dagger L_\ell \rangle \bigg) \nonumber\\
	&-\frac{i}{2}x\sum_\ell [H,L_\ell^\dagger L_\ell]\bigg\} \ket{\psi(t)}\Bigg|\Bigg|^2\,,
\label{eq:optimum_eq}
\end{align}
and we can obtain the optimum sequence $x(t)$ minimising Eq.~\eqref{eq:optimum_eq}.
The quantum circuit representation of unitary evolution can be naturally designed by the Trotterisation~\cite{smith2019} and the quantum circuit representation of non-unitary evolution can be implemented by MSSE.

\section*{Results}
We apply the DQS algorithm by the MSSE equation on the prototypical models in open quantum systems. Specifically, we deal with the one dimensional DTI model and the QCP model with open boundary condition. We firstly use the \textit{QASM simulator} and demonstrate that the DQS algorithm based on MSSE describes the exact dynamics with the large system size and the optimised quantum circuits are more accurate than original quantum circuits. Then we use the IBM Quantum devices including \textit{ibm\_hanoi} and the quantum emulator of \textit{ibmq\_tokyo}.
In all simulations, we perform the readout error mitigation method using the built-in \textsc{qiskit}~\cite{matthew_2022} library. We use 8192 measurement shots per data point and repeat three times.

\subsection*{Dissipative transverse Ising model}

\begin{figure}[!t]
	\includegraphics[width=0.8\columnwidth]{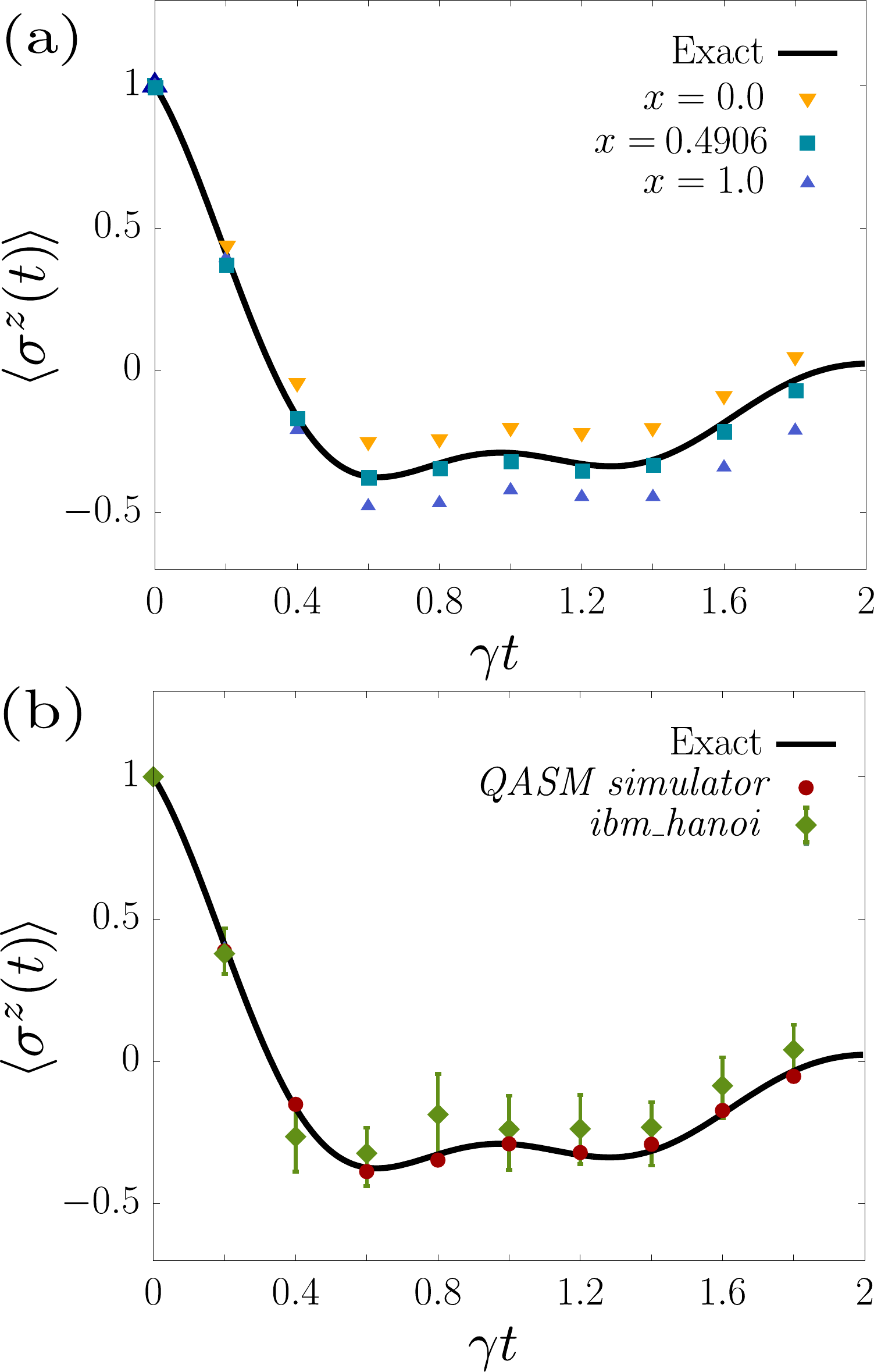}
	\caption{Results of the DTI model using the DQS algorithm by MSSE. Plot of the magnetisation $\langle \sigma^z(t) \rangle$ as a function of time $\gamma t$ for various optimised sequences $x(t)$: (a) constant sequences $x(t)=0$, $1$, and $0.4906$ (b) optimised sequence obtained by minimising Eq.~\eqref{eq:optimum_eq}. Symbols without error bars are obtained using the \textit{QASM simulator} and symbols with error bars (green symbols) are obtained using \textit{ibm\_hanoi}. The exact numerical result (black solid line) is obtained using \textsc{qutip} library. System size is taken $N=2$. 
	 \label{fig:fig1}}
\end{figure}

We consider the one-dimensional DTI model~\cite{ates2012,jin2018,rose2016,hu2013} of spin-$1/2$ particles. 
The Hamiltonian $H$ is expressed as
\begin{align}
    {H}=-J\sum_{\langle m,\ell \rangle} {\sigma}^z_{\ell}{\sigma}^z_{m}+\Delta\sum_\ell {\sigma}^x_{\ell}\,,
\label{eq:HS_DTI}
\end{align}
where $J$ represents the strength of ferromagnetic interaction of Ising spins in $z$ direction. The summation runs for nearest pair $\langle m,l \rangle$ of spins. $\Delta$ represents the strength of a transverse field and $\ell$ is spin index running from $\ell=1,\cdots N$. The decay Lindblad operators are given by
\begin{align*}
L_\ell = \sqrt{\gamma}\sigma_\ell^-\,,
\end{align*}
with the decay rate $\gamma$.

Note that the dissipative phase transition does not occur in one dimensional DTI model~\cite{marcuzzi2014}. Thus, the simulation with large system size is not necessary, so in this case we deal with the small system size using \textit{ibm\_hanoi} to test the quantum algorithm based on the MSSE. The initial state is fully up spins and parameters are fixed at $J=1$, $\Delta=1$, and $\gamma=0.5$. 

We first compare the constant sequence with the optimised sequence of $x(t)$ obtained by minimising Eq.~\eqref{eq:optimum_eq} for fixed $\gamma\Delta t=0.2$. In Fig.~\ref{fig:fig1}(a), we plot the dynamics of magnetisation for different constant values of $x(t)=0$, $1$, and $0.4906$. We analytically check that $x(0)$ converges to $0.5$ as the system size increases, i.e., $\lim_{N\to\infty}x(0)=0.5$. 
For $x(t)=0$ and $1$, the data deviate from the exact results. Practically, when $x(t)$ is taken from the initial value $x(0)=0.4906$, we observe that the data are more optimised than $x(t)=0$ and $1$. In Fig.~\ref{fig:fig1}(b), we plot the dynamics of magnetisation with optimised sequence $x(t)$ using the noiseless \textit{QASM simulator} and \textit{ibm\_hanoi}. We observe that the data with the optimised sequence show more accurate results than those with constant sequences. 
When we use \textit{ibm\_hanoi} to simulate the DTI model, the results show that it qualitatively follows the trend of the magnetisation; however, it is hard to describe the oscillation behaviour. Moreover, it eventually converges to 0 at large time limit due to the randomly sample state. Those discrepancies in the DQS algorithm is discussed by the noise model in the following section.

\begin{figure*}[!t]
	\includegraphics[width=2.1\columnwidth]{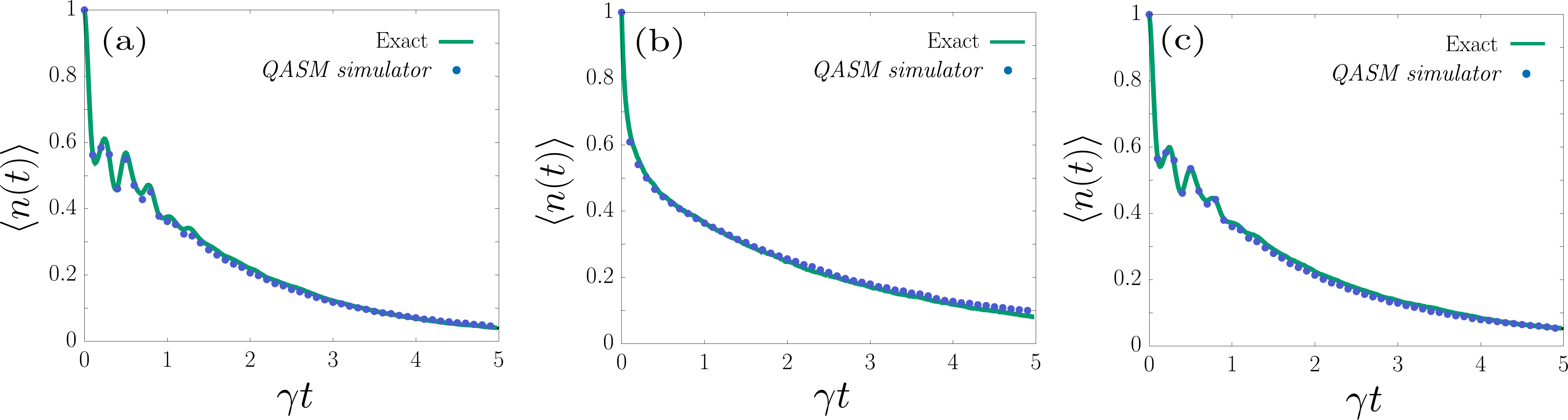}
	\caption{Results of the QCP model using the DQS algorithm by MSSE with the noiseless \textit{QASM simulator} simulator. Plot of $n(t)$ as a function of $t$ for the QCP model at the various parameter space $(\kappa, \omega)$: (a) $(0.0, 6.04)$, (b) $(6.0, 0.0)$, and (c) $(0.2,6.0)$. The system size is taken $N=4$. Blue dots are obtained by Trotterisation for the MSSE with the \textit{QASM simulator} with optimised sequence $x(t)$. The classical numerical result (green line) is obtained using \textsc{qutip} library.  	
	\label{fig:fig2}}
\end{figure*}

\begin{figure}[!t]
	\includegraphics[width=0.8\columnwidth]{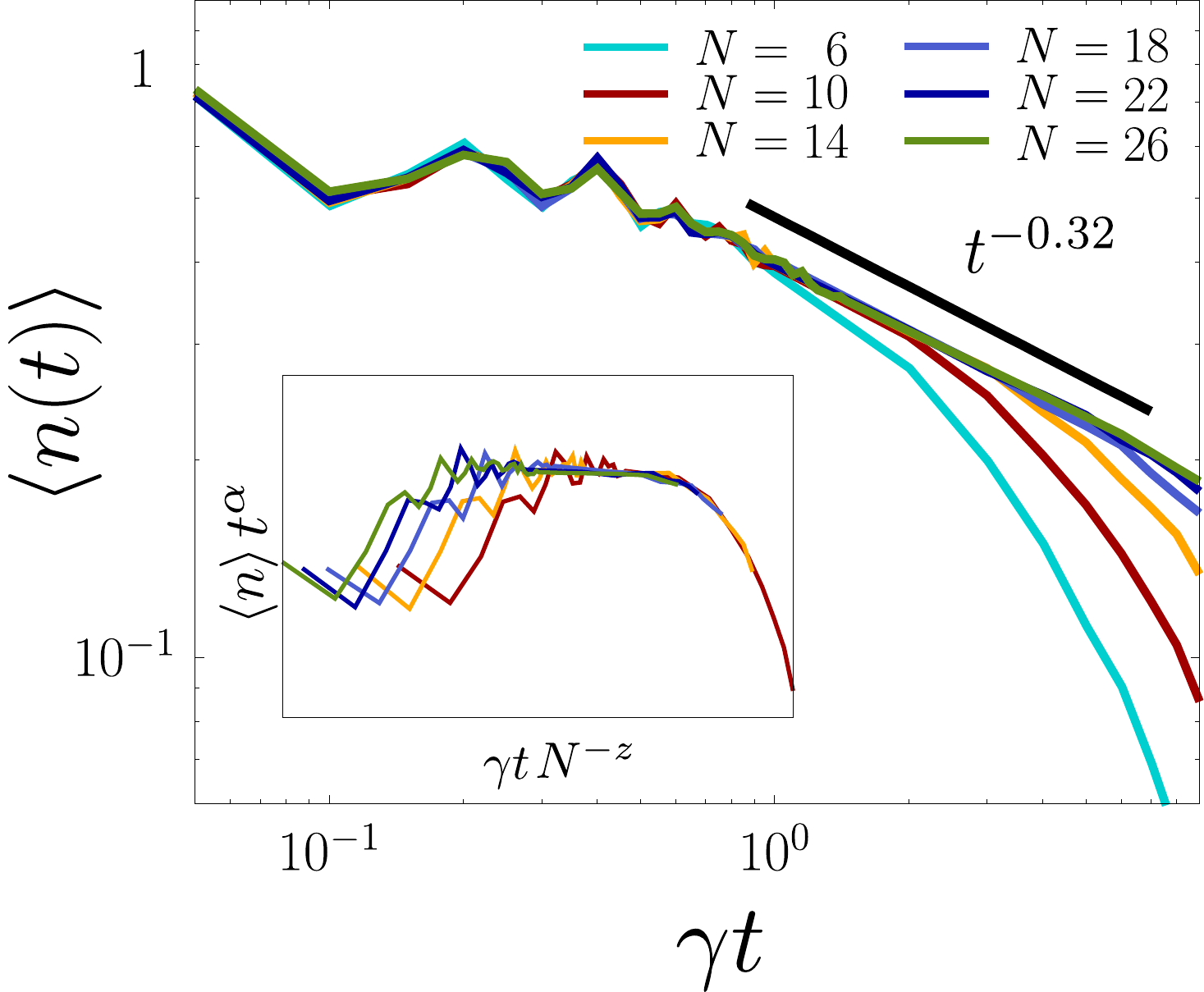}
	\caption{Estimates of the critical exponents $\alpha$ of the QCP model using the DQS algorithm by MSSE with the noiseless \textit{QASM simulator}. Plot of $n(t)$ as a function of $t$ on double-logarithmic scale for different system sizes at the critical point $(\kappa,\omega_c)=(0,6.04)$. The expected slope $n(t)\sim t^{-\alpha}$ with $\alpha=0.32$ is shown in black solid line with an arbitrary displacement for illustration. Inset: Scaling plot of $n(t)t^{\alpha}$ versus $tN^{-z}$. The data are well collapsed onto a single curve with $\alpha=0.32$ and $z=1.55$.
	 \label{fig:fig3}}
\end{figure}
\subsection*{Quantum contact process}

We consider a one-dimensional spin chain, where each site can be in an active (up-spin, $\left|\uparrow\right>$) or inactive (down-spin, $\left|\downarrow\right>$) state. 
The QCP model~\cite{marcuzzi2016, jo2019, gillman2021, gillman2022} consists of two coherent and three incoherent processes:
\begin{enumerate}
\item[(i)] Coherent branching or coagulation: This process is driven by a Hamiltonian, which satisfies the condition $\left<\mu\right| H \left|\nu\right> \neq 0$ such that coherent branching ($\left|\mu\right> = \left|\uparrow\downarrow\right>$ and $\left|\nu\right> = \left|\uparrow\uparrow\right>$), or coherent coagulation ($\left|\mu\right> = \left|\uparrow\uparrow\right>$ and $\left|\nu\right> = \left|\uparrow\downarrow\right>$), with rate $\omega$.
\item[(ii)] Decay: An active site spontaneously decays its state to an inactive state, which is denoted as $\left|\uparrow\right> \rightarrow \left|\downarrow\right>$, with rate $\gamma$.
\item[(iii)] Incoherent branching or coagulation: An active site incoherently activates (branching, $\left|\uparrow\downarrow\right> \rightarrow \left|\uparrow\uparrow\right>$) or inactivates (coagulation, $\left|\uparrow\uparrow\right> \rightarrow \left|\uparrow\downarrow\right>$) an inactive particle at the nearest-neighbour site at a rate $\kappa$.
\end{enumerate}

The coherent process (i) is described by the Hamiltonian, which is given by
\begin{align}
	H= \omega \sum_{\langle m, \ell \rangle} \left[n_{m}\left({\sigma}^+_\ell+{\sigma}^-_\ell \right)  \right] \,,
\label{eq:HS_QCP}
\end{align}
where $\langle m,\ell \rangle$ indicates $m$ and $\ell$ are the nearest neighbour.

Also, the Lindblad operators of decay (ii), branching and coagulation (iii) are expressed as
\begin{align}
	\label{eq:ld_QCP}
	{L}_{\ell}^{(d)} &= \sqrt{\gamma} {\sigma}_{\ell}^{-} \,, \quad
	{L}_{m\ell}^{(b)} &= \sqrt{\kappa} n_{m}{\sigma}^+_{\ell} \,, \quad
	{L}_{m\ell}^{(c)} &= \sqrt{\kappa} n_{m}{\sigma}^-_{\ell} \,,
\end{align}
respectively. The composite operators ${L}_{m\ell}^{(b, c)}$ with $\ell \neq m$ indicate that the active state at site $m$ activates or deactivates the state at $\ell$, representing the branching and coagulation processes. Instead, ${L}_{\ell}^{(d)}$ in Eq.~\eqref{eq:ld_QCP} denotes the decay dynamics of the active state at $\ell$. 

When $\kappa$ and $\omega$ are small compared to $\gamma$, inactive states become more abundant with time, and the system eventually falls into an absorbing state in which it is fully occupied by inactive states. By contrast, when $\kappa$ or $\omega$ is large compared to $\gamma$, the system remains in the active state, which is the nonequilibrium steady state with a finite density of active states. Thus, the QCP model exhibits a nonequilibrium phase transition from an active to an absorbing state.
Note that the QCP model with $\omega=0$ reduces to the classical contact process, which belongs to the classical directed percolation universality class at the critical point~\cite{jo2020}.

In the context of the universality class, one dimensional case is particularly interesting because the critical behaviour of the QCP model exhibits novel critical phenomena~\cite{carollo2019,jo2021}. Specifically, the dynamic critical exponent $\alpha$ changes from 0.32 at the quantum critical point $(\kappa,\omega_c)=(0,6.04)$ to 0.16 at the classical critical point. From now on, we obtain the critical exponent by measuring power-law behaviour based on the quantum circuits representation in the Methods. 

In Fig.~\ref{fig:fig2}, we perform the simulation for the various parameter space starting from the fully up spins at $\gamma=1$. The optimised sequence of $x(t)$ is obtained by minimising Eq.~\eqref{eq:optimum_eq}. We measure the up-spin density $n(t)$ of active sites at time $t$, which is formulated as $\langle n(t)\rangle=(\sum_{\ell}\text{Tr}[{\rho}(t){n}_{\ell}])/N$. In Fig.~\ref{fig:fig2}(a), we consider the quantum region $(\kappa, \omega_c)=(0.0, 6.04)$, where the classical branching and coagulation processes are absent. In Fig.~\ref{fig:fig2}(b), on the contrary, we plot the classical region $(\kappa,\omega)=(6.0, 0.0)$, where the quantum branching and coagulation processes are absent. In Fig.~\ref{fig:fig2}(c), we consider the intermediate region $(\kappa, \omega)=(0.2, 6.0)$, where the all processes are present. The results of quantum circuit representation using the \textit{QASM simulator} are consistent with classical exact results for various parameter spaces.

At the quantum critical point $(\kappa, \omega_c)=(0.0, 6.04)$, we find that $\langle n(t)\rangle$ exhibits power-law decay as $\langle n(t)\rangle \sim t^{-\alpha}$ with exponent $\alpha=0.32\pm 0.01$, as shown in Fig.~\ref{fig:fig3}. We note that the optimised sequence is taken as the initial value $x(t)=x(0)$. As system size increases, the slope of power-law can be precisely measured. Based on the critical exponent value, we perform the finite size scaling in the inset of Fig.~\ref{fig:fig3}. It shows data points collapse well onto a single curve for $\alpha=0.32\pm 0.01$ and $z=1.55\pm 0.03$, which is consistent with that of the classical computer~\cite{jo2021}. Note that the speed of the quantum algorithm by the \textit{QASM simulator} is more efficient so that we are able to the system size $N=26$ (see Fig.~\ref{fig:fig3}), which is challenging for a classical algorithm. At the system size, the power-law region is long enough to measure the slope for $\gamma t\in [1, 8]$.

\begin{figure}[!t]
	\includegraphics[width=0.8\columnwidth]{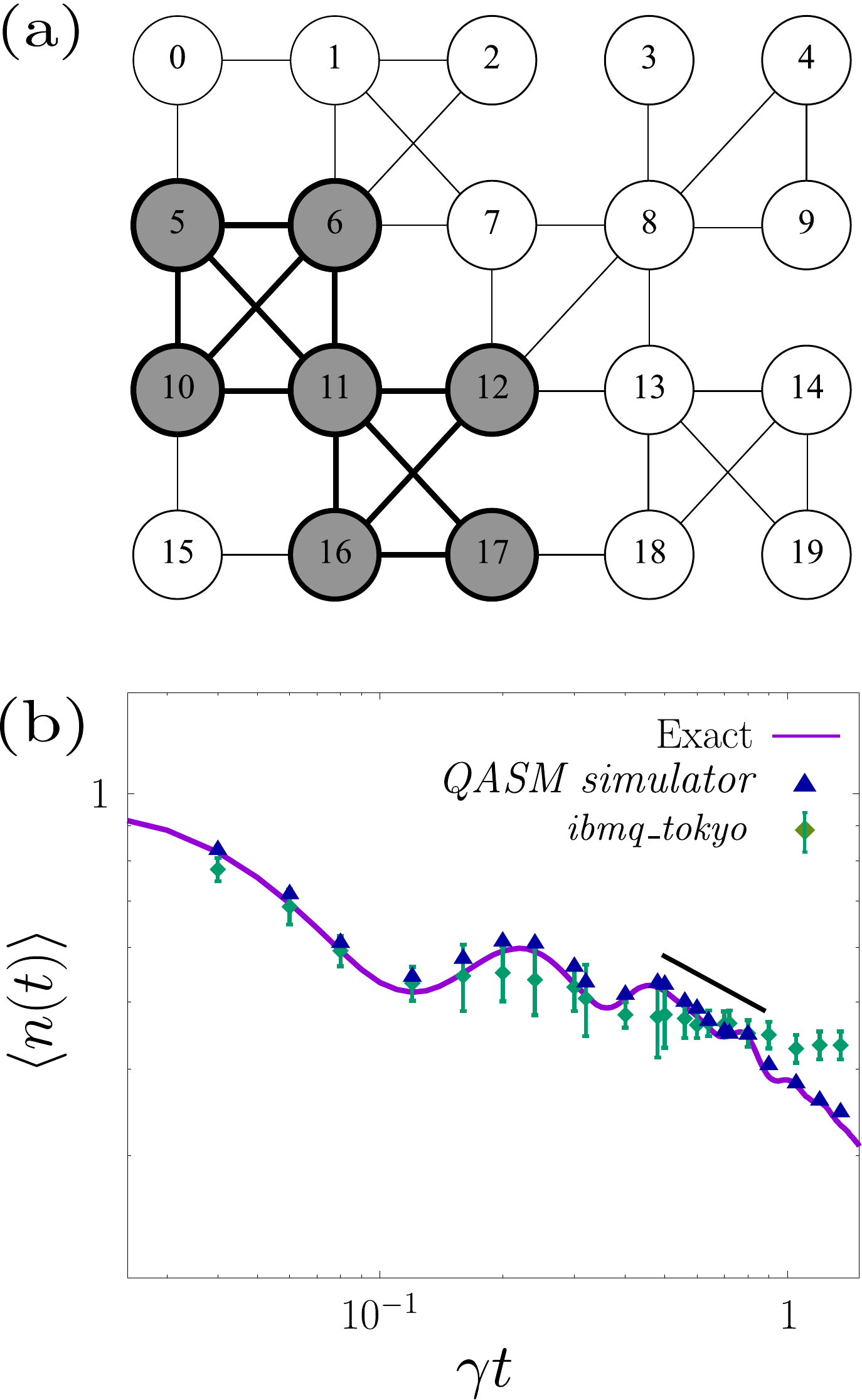}
	\caption{Results of the QCP model using the IBM Quantum device. (a) Topology of \textit{ibmq\_tokyo}. Grey qubits indicate chosen qubits. For $N=6$, we choose the qubit 11 as the ancilla qubit and remaining qubits as the system qubits. (b) Plot of the observable $n(t)$ as a function of time on double-logarithmic scale using the \textit{QASM simulator} and quantum emulator of \textit{ibmq\_tokyo}. The expected $n(t)\sim t^{-\alpha}$ slope is shown in black solid line with an arbitrary displacement for illustration. The quantum device follows the qualitative behaviour, but fails to predict the power law behaviour. 
	 \label{fig:fig4}}
\end{figure}

Next, we perform the simulation using the IBM Quantum device. In particular, we use the quantum emulator of \textit{ibmq\_tokyo} shown in Fig.~\ref{fig:fig4}(a), which is currently the best option due to the connectivity. The maximum system size of \textit{ibmq\_tokyo} is $N=6$ because the quantum algorithm based on MSSE requires a single ancilla qubit connected to the rest of system qubits. 
For $N=6$, we select the qubit 11 as the ancilla qubit and the other qubits $\{5, 6, 10, 12, 16, 17\}$ as the system qubits.
In Fig.~\ref{fig:fig4}(b), we plot the dynamics of the density $n(t)$ at the critical point of quantum region. Even though the data follow the trend of result of classical simulator, it is hard to measure the critical exponent because the system size is too small. In particular, oscillating behaviour near $\gamma t=2\times 10^{-1}$ deviates from the exact results. Moreover, the circuit depth increases linearly with the number of Trotter steps. This leads to more physical errors, which corrupt quantum simulations in noisy devices. As a result, for times $\gamma t > 1$, the magnetization (average of all spins) approaches zero because noise is so dominant that states belong to randomly sample states $\sigma^z(\gamma t>1)=0$ ($n(\gamma t>1)=0.5$). We remark that this random sample state was also observed in Ref.~\cite{smith2019} for the Trotterisation of the closed quantum system for $N=6$. 

\section*{Controlling gate error on noise model}

As the system size increases, the current quantum hardware does not provide sufficient performance to estimate the exact behaviour due to the noise. Thus, we discuss how much the noise needs to be reduced for the certain fidelity. To describe the noisy quantum device, we use the noise model in \textsc{qiskit} library, which enables us to control several noise factors. Typically, the noise of systems is characterised by single- and two-qubit gate errors, measurement readout error, and T1 and T2 times measuring how long each qubit can retain quantum information. 

To simplify the noise model, we make some assumptions. First, we take the gate errors modelled as a depolarising noise channel and control the two-qubit gate error $p$ and the single-qubit error $p/10$. This reflects for existing systems that the two-qubit gate errors is an order of magnitude higher than the 1-qubit gate error~\cite{saki2021,lubinski2021}. Moreover, among many different noise channels, the depolarising noise model appropriately describes the average noise in real devices containing many qubits and gates~\cite{urbanek2021,pathumsoot2020}. Second, we assume that the connectivity issue is resolved. In other words, there exists single qubit connected to the rest of the system qubits. Finally, we neglect the errors from thermal relaxation and readout.

\begin{figure}[!t]
	\includegraphics[width=0.8\columnwidth]{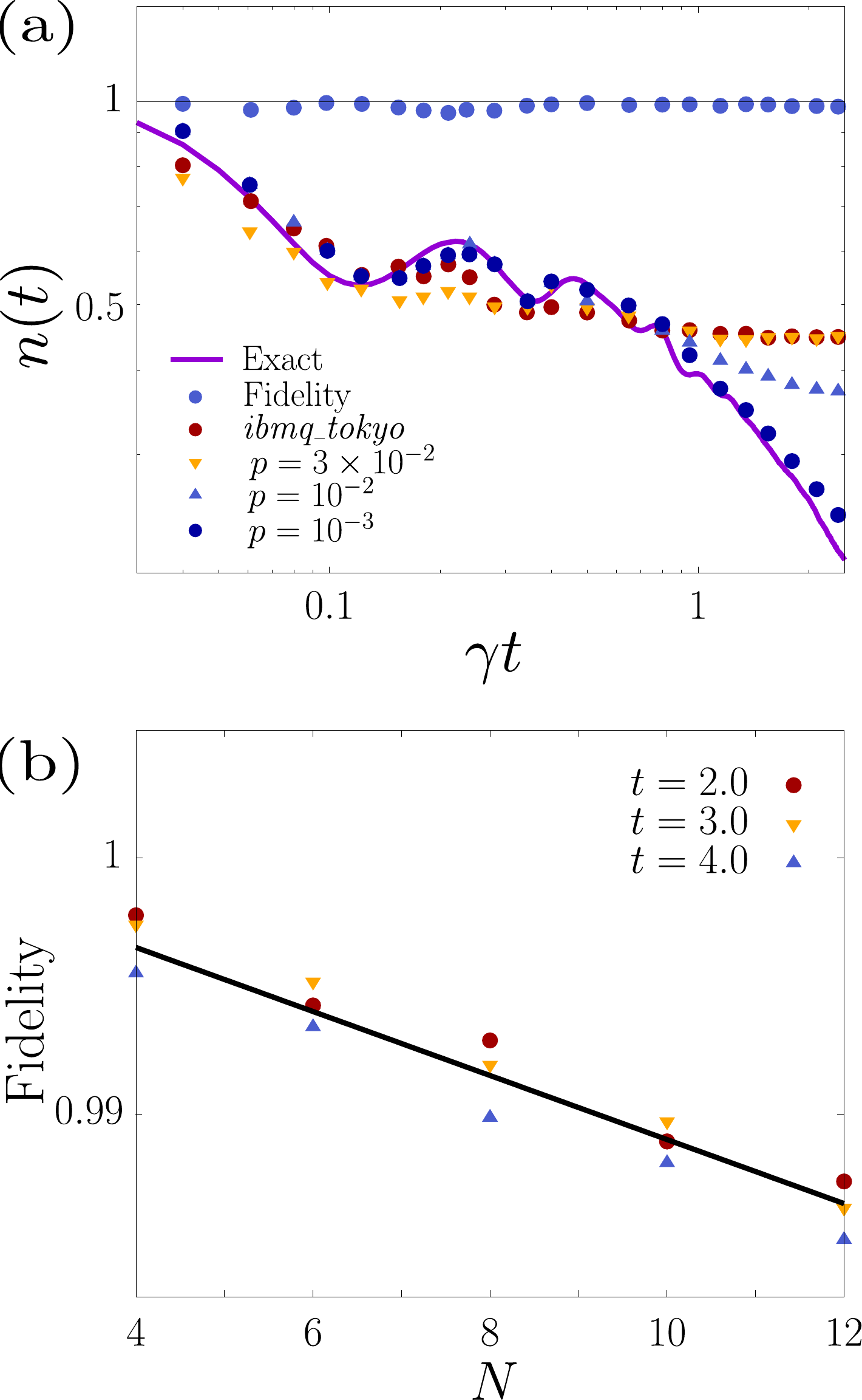}
	\caption{Results of the noise model. (a) Plot of the observable $n(t)$ as a function of time on double-logarithmic scale using the noise model with various gate noise $p$. At $p=0.03$ for \textit{ibmq\_tokyo}, the fidelity is nearly $1$ for time step $\gamma t<3$. The system size is taken $N=6$. (b) Plot of the fidelity as a function of the system size $N$ on semi-logarithmic scale at $p=10^{-3}$. Note that the fidelity exponentially decays with the system size. 
	 \label{fig:fig5}}
\end{figure}

To evaluate the performance of the two different machines, we use the state fidelity,
\begin{align}
F( \rho_1, \rho_2)= \text{Tr} \left[ \sqrt{\sqrt{\rho_1} \rho_2 \sqrt{\rho_1}} \right]^2 \,,
\end{align}
between the states $\rho_1$ and $\rho_2$. 
To test the validity of the above assumptions, we first perform the simulation on the noise model with \textit{ibmq\_tokyo}'s two-qubit error, i.e. $p=3\times 10^{-2}$. In Fig.~\ref{fig:fig5}(a), for the $N=6$ QCP model, the data from the noise model similarly describe the obtained experimental data from \textit{ibmq\_tokyo}. Moreover, they exhibit randomly sampled state for $\gamma t>1$. The fidelity between \textit{ibmq\_tokyo} and the noise model is nearly one for all time steps. As the gate error decreases, the noise model data converge to the exact data. At $p=10^{-3}$, the noise model successfully describes the oscillation region for $\gamma t<1$, power-law region for $\gamma t>1$. The current two-qubit gate error for most of IBM Quantum devices is approximately $\mathcal{O}(10^{-2})$, so the error should be reduced to be useful the method based on MSSE. 

At the fixed two-qubit error $p=10^{-3}$, a natural question arises as to whether the system size can achieve the certain fidelity. We set the threshold value of fidelity $F=0.95$, because at this fidelity we are able to estimate the the power-law behaviour accurately. In Fig.~\ref{fig:fig5}(b), we plot the fidelity as a function of the system size in semi-logarithmic scale at time steps where the data exhibit the power-law behaviour. The fidelity scales as the exponential function with the system size $N$. Based on the extrapolation, the data indicate that DQS algorithm based on the MSSE will be manageable up to the system size $N=40$ that is intractable by classical computer.

\section*{Summary and Conclusions}
We have investigated a DQS algorithm of the MSSE for time dynamics of open quantum systems. Analysing the leading error in MSSE, we find there is the optimum quantum circuit originating from the non-commutativity between Hamiltonian and Lindblad operators. The optimised quantum circuit representation enables to perform the simulation at a relatively large discretised time interval, which is useful for constrained resources of quantum simulation. Moreover, we use a single ancilla qubit supported by reset gates and mid-circuit measurements to implement non-unitary dynamics. Minimising the number of ancilla qubits by these gates enables to perform the simulation with the large system size, which is useful to determine the universality class of many-body systems. However, these gates are currently supported by IBM devices, so we have examined the DQS algorithm by the noiseless \textit{QASM simulator} and noisy IBM Quantum devices.

The DQS algorithm by the \textit{QASM simulator} enables to simulate at a relatively large system size comparable to the limits of the classical computer. Also the consuming time of DQS algorithm, which grows linearly with the system size and the number of Trotter steps, is fast due to the absence of the classical optimisation. We run the optimised quantum circuits and successfully obtain the critical exponent of the QCP model consistent with the classical computer with the large system size $N=26$. 

The goal of the DQS algorithm is to perform the simulation with the large system size beyond the classical algorithm. We have used noisy IBM Quantum devices, and the consistent data were obtained at the system size $N=6$; however, it is challenging to observe the power-law behaviour clearly due to the noise. In the DQS algorithm, the noise particularly hinders the oscillatory behaviour and long-time simulation.
To study the discrepancy, we investigate the noise model based on the depolarising channel and find that it is mainly due to the gate noise of real quantum devices. 
Based on the noise model, the extrapolation at the fixed gate errors $p=10^{-3}$, which is approximately a factor of 1 smaller than the current errors of IBM Quantum devices, suggests that the method based on the MSSE will be tolerable up to system size $N=40$ with the fidelity $F\approx 0.95$. This system size is intractable by the classical computers and thus DQS algorithm surpasses the classical algorithm. 


\begin{appendix}

\section*{Method}
\noindent\textbf{Lindblad jump}\\
From now on, all differentials $dt$ will be replaced with small but finite differences $\Delta t$. Let us consider the Lindblad operation represented in the square brackets of Eq.~\eqref{eq:MSSE}
\begin{align}
	\label{eq:jump_eq}
	|\psi(t+\Delta t)\rangle\otimes \ket{a}&=\Big( 1-\sum_\ell\frac{1}{2}{L}_\ell^\dagger {L}_\ell\Delta t\Big) |\psi(t)\rangle\otimes |0\rangle \nonumber\\
	&+\sum_\ell  L_\ell\sqrt{\Delta t}|\psi(t)\rangle\otimes |1\rangle
	\,.
\end{align}
	
Once the Lindblad operator is determined, a quantum circuit can be designed by Eq.~\eqref{eq:MSSE}. For instance, we show the quantum circuit representing the three Lindblad operators used in dissipative transverse Ising model and quantum contact process, ${L}_\ell=\sqrt{\gamma} \pauliminus{\ell}$, ${L}_{m\ell}=\sqrt{\kappa} \n{m}\pauliminus{\ell}$, and ${L}_{m\ell}=\sqrt{\kappa} \n{m}\pauliplus{\ell}$. 
Here, $n_m=\ket{\uparrow}\bra{\uparrow}_m$ is the up-spin number operator at site $m$. $\pauliplus{\ell}=\ket{\uparrow}\bra{\downarrow}_\ell$ and $\pauliminus{\ell}=\ket{\downarrow}\bra{\uparrow}_\ell$ are the spin-raising and -lowering operators at site $\ell$, respectively. 

Firstly, let us consider the decay operator ${L}_\ell=\sqrt{\gamma} \pauliminus{\ell}$, which was already implemented in the amplitude damping channel~\cite{nielsen00}. Eq.~\eqref{eq:MSSE} of this Lindblad operator for $\ket{\psi(t)}\otimes \ket{0} = (\alpha_0\ket{0}+\alpha_1\ket{1})\otimes\ket{0}$ becomes
\begin{align}
\left[ \alpha_0\ket{0} +\alpha_1 \left( 1-\frac{1}{2}\gamma \Delta t\right) \ket{1}\right] \otimes |0\rangle 
	+\alpha_1\sqrt{\gamma \Delta t}\ket{0}\otimes \ket{1}\,,
\end{align}
which was achieved by the following circuit until the dotted line
\begin{align}
\Qcircuit @C=1em @R=.7em 
{
     \lstick{\ket{\psi}} & \ctrl{1} \ar@{.}[]+<3.5em,1em>;[d]+<3.5em,-1em> & \targ & \qw & \qw & \\
     \lstick{\ket{0}} & \gate{R_y(\theta)} & \ctrl{-1} & \meter &\qw & & \lstick{\ket{0}} &\qw
}
\end{align}
This circuit until the dotted line maps an initial state to the state
\begin{align}
	\label{eq:circuit_decay}
	\left[ \alpha_0\ket{0} +\alpha_1 \cos\left(\frac{\theta}{2}\right)\ket{1}\right] \otimes |0\rangle 
	+\alpha_1\sin\left(\frac{\theta}{2}\right)\ket{0}\otimes \ket{1}
	\,.
\end{align}
If we set $\theta = 2\arcsin(\sqrt{\gamma \Delta t})$ with $\Delta t\ll 1$, Eq.~\eqref{eq:circuit_decay} becomes Eq.~\eqref{eq:jump_eq} describing decay process.

Next, we consider the composite Lindblad operator ${L}_{m\ell}=\sqrt{\kappa} \n{1}\pauliminus{2}$. This can be achieved by the following circuit until the dotted line
\begin{align}
\Qcircuit @C=1em @R=.7em {
      & \ctrl{2} \ar@{.}[]+<3.5em,0.5em>;[d]+<3.5em,-2.5em>& \qw & \qw & \qw &\\
     \lstick{\raisebox{1.4em}{$\ket{\psi}$\ }} & \ctrl{1} & \targ & \qw & \qw &\\
     \lstick{\ket{0}} & \gate{R_y(\theta)} & \ctrl{-1} & \meter &\qw & & \lstick{\ket{0}} &\qw
}
\end{align}
This circuit until the dotted line maps an initial state $\ket{\psi}\otimes \ket{0} = (\alpha_{00}\ket{00}+\alpha_{01}\ket{01}+\alpha_{10}\ket{10}+\alpha_{11}\ket{11})\otimes\ket{0}$ to the state
\begin{align}
	\label{eq:circuit_branching}
	&\left[ \alpha_{00}\ket{00}+\alpha_{01}\ket{01}+\alpha_{10}\ket{10}+\alpha_{11}\cos\left(\frac{\theta}{2}\right)\ket{11}
	\right] \otimes |0\rangle \nonumber\\
	&+\alpha_{11}\sin\left(\frac{\theta}{2}\right)\ket{10}\otimes \ket{1}
	\,.
\end{align}
If we set $\theta = 2\arcsin(\sqrt{\kappa \Delta t})$ with $\Delta t\ll 1$, Eq.~\eqref{eq:circuit_branching} becomes Eq.~\eqref{eq:jump_eq} describing coagulation process.

Likewise, ${L}_{m\ell}=\sqrt{\kappa} \n{1}\pauliplus{2}$ can be achieved by the following circuit until the dotted line
\begin{align}
\Qcircuit @C=1em @R=.7em {
      & \ctrl{2} \ar@{.}[]+<3.5em,0.5em>;[d]+<3.5em,-2.5em>& \qw & \qw & \qw &\\
     \lstick{\raisebox{1.4em}{$\ket{\psi}$\ }} & \ctrlo{1} & \targ & \qw & \qw &\\
     \lstick{\ket{0}} & \gate{R_y(\theta)} & \ctrl{-1} & \meter &\qw & & \lstick{\ket{0}} &\qw
}
\end{align}
This circuit until the dotted line maps an initial state $\ket{\psi}\otimes \ket{0} = (\alpha_{00}\ket{00}+\alpha_{01}\ket{01}+\alpha_{10}\ket{10}+\alpha_{11}\ket{11})\otimes\ket{0}$ to the state
\begin{align}
	\label{eq:circuit_coagulation}
	&\left[ \alpha_{00}\ket{00}+\alpha_{01}\ket{01}+\alpha_{10}\cos\left(\frac{\theta}{2}\right)\ket{10}+\alpha_{11}\ket{11}
	\right] \otimes |0\rangle \nonumber\\
	&+\alpha_{10}\sin\left(\frac{\theta}{2}\right)\ket{10}\otimes \ket{1}
	\,.
\end{align}
If we set $\theta = 2\arcsin(\sqrt{\kappa \Delta t})$ with $\Delta t\ll 1$, Eq.~\eqref{eq:circuit_coagulation} becomes Eq.~\eqref{eq:jump_eq} describing branching process.

$ $\\
\noindent\textbf{Hamiltonian evolution}\\
Following Ref.~\cite{smith2019}, we construct the quantum circuit representation of unitary Hamiltonian evolution.
Let us consider the unitary operator $U=e^{-i{H}\Delta t}$ where $ H=\sum_j A_j+\sum_j B_{j,j+1}$ and $A$ and $B$ are a single-qubit operator and a two-qubit operator, respectively. To construct the quantum circuit representing this unitary operator, we use the symmetric Trotterizaition:
\begin{align}
 U&=\prod_j e^{-i  A_j \frac{\Delta t}{2}}\prod_{j\in \rm{even}} e^{-i  B_{j,j+1}\frac{\Delta t}{2}}
 \prod_{j\in \rm{odd}} e^{-i  B_{j,j+1}{\Delta t}}
 \prod_{j\in \rm{even}} e^{-i  B_{j,j+1}\frac{\Delta t}{2}}\nonumber\\
 &\times\prod_j e^{-i A_j\frac{\Delta t}{2}}+\mathcal{O}(\Delta t^3)\,.
 \label{eq:trotter}
\end{align}
For $N=4$, a quantum circuit expressing Eq.~\eqref{eq:trotter} is
\begin{align}
 \Qcircuit @C=1em @R=.7em {
    & \gate{e^{-i  A_0 \frac{\Delta t}{2}}} 
    & \multigate{1}{e^{-i  B_{0,1}\frac{\Delta t}{2}}} 
    & \qw
    & \multigate{1}{e^{-i  B_{0,1}\frac{\Delta t}{2}}} 
    & \gate{e^{-i  A_0 \frac{\Delta t}{2}}} 
    & \qw \\
    & \gate{e^{-i  A_1 \frac{\Delta t}{2}}} 
    & \ghost{e^{-i  B_{0,1}\frac{\Delta t}{2}}} 
    & \multigate{1}{e^{-i  B_{1,2}{\Delta t}}}        
    & \ghost{e^{-i  B_{0,1}\frac{\Delta t}{2}}} 
    & \gate{e^{-i  A_1 \frac{\Delta t}{2}}}
    & \qw \\
    & \gate{e^{-i  A_2 \frac{\Delta t}{2}}} 
    & \multigate{1}{e^{-i  B_{2,3}\frac{\Delta t}{2}}} 
    & \ghost{e^{-i  B_{1,2}{\Delta t}}} 
    & \multigate{1}{e^{-i  B_{2,3}\frac{\Delta t}{2}}} 
    & \gate{e^{-i  A_2 \frac{\Delta t}{2}}} 
    & \qw \\
    & \gate{e^{-i  A_3 \frac{\Delta t}{2}}} 
    & \ghost{e^{-i  B_{2,3}\frac{\Delta t}{2}}} 
    & \qw
    & \ghost{e^{-i  B_{2,3}\frac{\Delta t}{2}}} 
    & \gate{e^{-i  A_3 \frac{\Delta t}{2}}}
    & \qw 
  }
  \label{eq:trotter_circuit}
\end{align}
In the above circuit, decomposition of two-qubit gate containing $B$ operator is non-trivial. By using the Cartan's KAK decomposition, any two-qubit unitary operator SU(4) can be decomposed as $U=(A_1\otimes A_2)U_D(A_3\otimes A_4)$ where $A_1, A_2, A_3$, and $A_4$ are single-qubit gates. Also $U_D=e^{-i(k_1\paulix{}\otimes\paulix{}+k_2\pauliy{}\otimes\pauliy{}+k_3\pauliz{}\otimes\pauliz{})}$ where $k_1, k_2$, and $k_3$ are real numbers. To decompose a two-qubit gate to the Kronecker product of two single-qubit gates, we use the Magic gate defined as
\begin{align}
\Qcircuit @C=1em @R=.7em {
&
\raisebox{-1.2em}{$M=\frac{1}{\sqrt{2}}\left(\begin{array}{cccc} 
1 & i & 0 & 0\\ 
0 & 0 & i & 1\\ 
0 & 0 & i & -1\\ 
1 & -i & 0 & 0\\ 
\end{array}\right)=$} 
&&&&&&& \gate{S} & \qw & \targ & \qw & \\
&&&&&&&& \gate{S} & \gate{H} & \ctrl{-1} &\qw\\
     }
   \label{eq:magic_gate}
\end{align}
where $H$ is the Hadamard gate defined as
\begin{align}
H=\frac{1}{\sqrt{2}}\left(\begin{array}{cc} 
1 & 1\\ 
1 & -1\\ 
\end{array}\right)\,.
\end{align}

For instance, let us construct the quantum circuit of $U=e^{-i\omega\Delta t (\pauliz{1}\paulix{2}+\paulix{1}\pauliz{2})}$ with the real coefficient $\omega$. Applying the KAK decomposition, we have
\begin{align}
U=(I\otimes H)e^{-i\omega\Delta t (\pauliz{1}\pauliz{2}+\paulix{1}\paulix{2})}(I\otimes H)\,.
\label{eq:KAK}
\end{align} 
Using Eq.~\eqref{eq:magic_gate} with the relations $M^\dagger (\paulix{1}\otimes\paulix{2})M=I\otimes \pauliz{2}$ and $M^\dagger (\pauliz{1}\otimes\pauliz{2})M=\pauliz{1}\otimes I$, Eq.~\eqref{eq:KAK} can be written as
\begin{align}
U=(I\otimes H)M(e^{-i\omega\Delta t \pauliz{1}}\otimes e^{-i\omega\Delta t \pauliz{2}})M^{\dagger}(I\otimes H)\,.
\end{align} 
This is represented by the quantum circuit
\begin{align}
\Qcircuit @C=1em @R=.7em {
&\qw&\targ&\qw&\gate{S^{\dagger}}&\gate{R_z(-2\omega\Delta t)}& \gate{S} & \qw & \targ & \qw & \qw\\
&\gate{H}&\ctrl{-1}&\gate{H}&\gate{S^{\dagger}}&\gate{R_z(-2\omega\Delta t)} & \gate{S} & \gate{H} & \ctrl{-1} &\gate{H} & \qw
\gategroup{1}{3}{2}{5}{.7em}{--}
\gategroup{1}{7}{2}{9}{.7em}{--}\\ \\
& & & M^\dagger & &&&M \\
     }
\end{align}
Moreover, the circuit can be further simplified using $[S,R_z(\theta)]=0$ which is expressed as
\begin{align}
\Qcircuit @C=1em @R=.7em {
&\qw&\targ&\gate{R_z(-2\omega\Delta t)}&\targ & \qw & \qw\\
&\gate{H}&\ctrl{-1}&\gate{R_x(-2\omega\Delta t)} & \ctrl{-1} &\gate{H} & \qw
     }
\end{align}
$ $\\
\noindent\textbf{Collecting data from quantum devices}\\
Due to the connectivity issue, two-qubit errors of IBM Quantum devices are unavoidable compared to the fully-connected device.
To minimise the CNOT gate for a given quantum circuit, we should note that the transpiler of \textsc{qiskit} varies the number of CNOT gates for the reproduction of the same experiment. Thus, it would be helpful to check minimum number of CNOT gates is inserted after logical-to-physical mapping. The other option is to use transpiler of \textsc{pytket}~\cite{sivarajah2020}.

Once we collect the data, we select the reliable data below the certain number of CNOT gates and the small Trotterisation time $\Delta t$. We use the the following steps:
\begin{enumerate}
\item Find the best qubits that gives the the long T2 (dephasing) time, low measurement error, and low CNOT errors (See Ref.~\cite{smith2019} how to choose the best qubits).
\item Once the best qubits are determined by step 1, determine minimum $\Delta t^*$ that the \textit{QASM simulator} describes the actual classical simulation.
\item For $\Delta t<\Delta t^*$, we perform the simulations and collect the data.
\item Among those data, we choose the data whose CNOT gate is less than the certain number for given CNOT error.
\item Perform the readout error mitigation.		
\end{enumerate}

\end{appendix}

\begin{acknowledgments}
We acknowledge discussions with Christopher Self. This research was supported by the quantum computing technology development program of the NRF funded by the Ministry of Science and ICT, No.~2021M3H3A103657312 (MJ).
We acknowledge the a Samsung GRC project and the UK Hub in Quantum Computing and Simulation, part of the UK National Quantum Technologies Programme with funding from UKRI EPSRC grant EP/T001062/1.
We acknowledge the use of IBM Quantum services for this work. The views expressed are those of the authors, and do not reflect the official policy or position of IBM or the IBM Quantum team.
\end{acknowledgments}

\section*{COMPETING INTERESTS}
The authors declare no competing interests.
\section*{Contributions}
All authors contributed to the design and implementation of the research, to the analysis of the results, and to the writing of the manuscript.
	
%

\end{document}